\documentclass[prb,twocolumn,showpacs,preprintnumbers,amsmath,amssymb,superscriptaddress,floatfix]{revtex4-2}

\usepackage{amsmath}    
\usepackage{amssymb,environ}
\usepackage{graphicx}   
\usepackage{verbatim}   
\usepackage{color}      
\usepackage{hyperref}   
\usepackage{breakurl}
\usepackage[normalem]{ulem}
\usepackage{natbib}
\usepackage{enumitem}
\usepackage{multirow}

\usepackage{epsfig}
\usepackage{textcomp}
\usepackage{wasysym}
\usepackage{array}
\usepackage{color}

\begin{document}

\newcommand{\ev}[0]{\mathbf{e}}
\newcommand{\cv}[0]{\mathbf{c}}
\newcommand{\fv}[0]{\mathbf{f}}
\newcommand{\Rv}[0]{\mathbf{R}}
\newcommand{\Tr}[0]{\mathrm{Tr}}
\newcommand{\ud}[0]{\uparrow\downarrow}
\newcommand{\Uv}[0]{\mathbf{U}}
\newcommand{\Iv}[0]{\mathbf{I}}
\newcommand{\Hv}[0]{\mathbf{H}}

\setlength{\jot}{2mm}

\newcommand{\jav}[1]{{\color{red}#1}}
\newcommand{\ds}[1]{{\color{blue}#1}}

\title{Infinite temperature spin dynamics in the asymmetric Hatsugai-Kohmoto model}

\author{\'Ad\'am B\'acsi}
\email{bacsi.adam@sze.hu}
\affiliation{Department of Mathematics and Computational Sciences, Sz\'echenyi Istv\'an University, 9026 Gy\H or, Hungary}

\author{Doru Sticlet}
\affiliation{National Institute for R\&D of Isotopic and Molecular Technologies, 67-103 Donat, 400293 Cluj-Napoca, Romania}

\author{C\u at\u alin Pa\c scu Moca}
\affiliation{Department of Theoretical Physics, Institute of Physics, Budapest University of Technology and Economics, M\H uegyetem rkp. 3., H-1111 Budapest, Hungary}
\affiliation{Department of Physics, University of Oradea,  410087, Oradea, Romania}

\author{Bal\'azs D\'ora}
\email{dora.balazs@ttk.bme.hu}
\affiliation{Department of Theoretical Physics, Institute of Physics, Budapest University of Technology and Economics, M\H uegyetem rkp. 3., H-1111 Budapest, Hungary}

\begin{abstract}
We focus on the infinite temperature dynamical spin structure factor of the asymmetric Hatsugai-Kohmoto model, the relative of the asymmetric Hubbard model.
 It is characterized by distinct single particle energies  for the
two spin species, which interact with each other through a contact interaction in momentum space. 
We evaluate its spin structure factor exactly and follow the evolution of its excitation spectrum for all fillings and interactions, identify signatures of the Mott transition and fingerprints of the asymmetric hoppings.
The longitudinal spin structure factor exhibits sound like and interaction induced gapped excitations, whose number gets doubled in the presence of hopping asymmetry.
The transverse response displays the competition of interaction and asymmetry induced gaps and results in a quadratic excitation branch at their transition.
The complete asymmetric case features momentum-independent dynamical structure factor, characteristic to transitions involving a flat band.


\end{abstract}

\maketitle

\section{Introduction}

Understanding universal features in the equilibrium and out-of-equilibrium dynamics of strongly interacting quantum systems represents a major challenge\cite{polkovnikovrmp,dziarmagareview}.
For classical or quantum spins\cite{nowak,blundell}, this is often probed by the dynamical spin structure factor, which reveals critical phenomena, spin diffusion, and the nature of collective magnetic excitations\cite{coldea}.
All of these are essential for understanding both classical and quantum materials, with applications in areas like magnetism, quantum materials, spintronics and quantum technologies.
In spite of its importance, the calculation of the dynamical properties in strongly correlated systems proves to be notoriously difficult.

Particularly interesting is infinite temperature spin dynamics\cite{niemijer,sur,gerling,katsura} which provides insights into the behavior of quantum and classical systems in highly disordered states. There, a system is maximally 
entropic, every possible state is equally likely, offering a useful benchmark for understanding how systems behave when they are far from equilibrium.
For example, the spin dynamics in a one-dimensional Heisenberg chain at infinite temperature exhibit anomalous correlations\cite{ljubotina,sanchez,gopalakrishnan}, while
infinite temperature Floquet systems can display prethermal regimes\cite{luitz}.

Recently, a distinct class of strongly correlated systems, the Hatsugai-Kohmoto model, was reintroduced\cite{hatsugai1992,Sun2024,zaanen,nesselrodt,zhao2022,lidsky,zhu2021,zhao2025,zhao2022,skolimowski} in condensed matter physics.
It describes electrons with long range interactions, such that the center of mass of the interacting electrons is conserved. While its Hamiltonian looks shockingly simple at first glance, 
it displays a rich phenomenology,
including non-Fermi liquid ground state, 
Mott transition with a possible flavour of topology\cite{morimoto}.
When supplemented with superconductivity\cite{Phillips2020}, it accounts for some prominent features of high temperature superconductors as well.

Motivated by these considerations, the infinite-temperature spin dynamics of the Hatsugai-Kohmoto 
model emerges as both conceptually rich and potentially relevant for understanding systems 
exhibiting exotic quantum phenomena. In this work, we focus on the asymmetric, or mass-imbalanced, 
variant of the Hatsugai-Kohmoto model, which continuously interpolates between the 
spin-rotationally invariant limit and the Falicov-Kimball regime \cite{falicov}. Using the model’s 
exact solvability, we derive and analytically evaluate the dynamical spin structure factor at 
infinite temperature in one dimension. Notably, many of our findings extend naturally to 
higher-dimensional settings.


We find that the longitudinal spin structure factor contains six sound modes in general away from the strict SU(2) symmetric limit. 
The transverse component displays more exotic behaviour and highlights the  competition of interaction and asymmetry induced gaps, which result 
in a quadratic low energy excitation branch at their transition.
We expect our results to describe at least qualitatively the infinite temperature spin dynamics of strongly correlated itinerant electron systems, including the interaction 
induced shift of spectral weight. 
The dynamical spin structure factor can be probed using techniques such as inelastic neutron scattering, resonant inelastic X-ray scattering, electron spin resonance, light absorption experiments, and quantum simulations\cite{baez}.






\section{The asymmetric Hatsugai-Kohmoto model}
The asymmetric or mass imbalanced Hatsugai-Kohmoto model \cite{hatsugai1992,Phillips2020,zhao2022} is defined by the Hamiltonian
\begin{gather}
H_0=\sum_k \Big(\varepsilon_\uparrow(k) n_{k\uparrow} + \varepsilon_\downarrow(k)n_{k\downarrow} + U n_{k\uparrow} n_{k\downarrow}\Big),
\label{ahk}
\end{gather}
where $n_{k\sigma} = c_{k\sigma}^+ c_{k\sigma}$ is the occupation number operator of the electrons with momentum $k$ and spin $\sigma$, $\varepsilon_\sigma(k)$ is the non-interacting band structure, which also contains the chemical potential $\mu$ and $U$ stands for
the Hatsugai-Kohmoto interaction.
Therefore, the present model allows for studying the asymmetric Hatsugai-Kohmoto model, similarly to the asymmetric Hubbard model\cite{Farkasovsky,maska,wang2007,heitmann,jin2015}. 
It interpolates between the SU(2) invariant case with 
$\varepsilon_\uparrow(k)=\varepsilon_\downarrow(k)$ and the Falicov-Kimball limit\cite{falicov,zlatic,freericks,Li2019} with $\varepsilon_\downarrow(k)=0$,
representing a simple model for a semiconductor-metal transition.
 For each momentum $k$, the Hamiltonian is diagonalizable in the four-dimensional many-body Hilbert space spanned by $|0\rangle$, $|\!\uparrow\rangle$, $|\!\downarrow\rangle$ and $|\!\ud\rangle$ denoting the states 
with zero particle, one spin up particle, one spin down particle and two particles, respectively. The eigenvalues are 
$0$ for the empty state, $\varepsilon_\sigma(k)$ for singly occupied state with either a spin up or a spin down particle, and 
 $\varepsilon_\uparrow(k)+ \varepsilon_\downarrow(k) + U$ for the doubly occupied case. 

The ground state phase diagram of the Hatsugai-Kohmoto model is well studied in the SU(2) invariant case\cite{Phillips2020}, and consists of non-Fermi liquid phases and a Mott insulating state at half-filling, irrespective of the spatial dimensionality. 
On the other hand, 
the Falicov-Kimball limit has received hardly any attention, therefore we discuss its ground state configuration briefly.
Since the total spin is also conserved in Eq.~\eqref{ahk}, the ground state structure also depends on it. For simplicity, we
focus on the sector with zero net magnetization. 
In the symmetric case, the ground state realizes a non-Fermi liquid for all filling but half, when a Mott transition occurs upon increasing the interaction\cite{Phillips2020}. 
In this case, the single particle band $\varepsilon_\uparrow(k)=\varepsilon_\downarrow(k)$ is singly occupied for
all $k$ with either up or down spin electrons. Due to the spin degree of freedom, this ground state manifold is highly degenerate and is protected by an excitation gap.
This realizes the minimal setting for an integrable, minimally many-body, SU(2) invariant model.

For the asymmetric case, away from half-filling, there are always low energy excitations present in the system, 
therefore it realizes a non-Fermi liquid state, similarly to its SU(2) invariant counterpart. At half filling, 
by taking a dispersion $\varepsilon_\sigma(k)=-t_\sigma\cos(ka)$ with $t_\uparrow\geq t_\downarrow\geq 0$, 
the ground state becomes unique and non-degenerate. All $\uparrow$ states are filled for $|k|<\pi/2$ 
while all $\downarrow$ state occupy the region  $|k|>\pi/2$ for any $U>U_c=2t_\downarrow$.
Creating an excitation costs a finite energy, therefore this state is also a Mott insulator, but 
the corresponding spin configuration is different from that of the SU(2) invariant case.
The $U_c$ vanishes for the fully asymmetric case with a flat band with $t_\downarrow=0$, the Mott transition takes place for any infinitesimal value of interaction.
Thus, momentum space phase separation occurs between the two spin states, analogously to the real space Falicov-Kimball model.
Very similar considerations apply also to the higher dimensional cases as well.

\section{High temperature Green's function}
Having discussed the ground state properties, we focus on the infinite temperature properties.
The single particle Green's function of the electrons is obtained for Eq.~\eqref{ahk} as
\begin{gather}
G_\sigma(k,\omega_n)=\frac{1-\langle n_{k~ -\sigma}\rangle}{i\omega_n-\varepsilon_\sigma(k)}+\frac{\langle n_{k~ -\sigma}\rangle}{i\omega_n-\varepsilon_\sigma(k)-U},
\end{gather}
with 
 $\omega_n=(2n+1)\pi T$ the fermionic Matsubara frequency with $n$ integer and $T$ the temperature. The denominators contain the energy of the lower and upper Hubbard bands\cite{arovas}, 
$\varepsilon_\sigma(k)$ and $\varepsilon_\sigma(k)+U$, respectively.
The occupation number is
\begin{gather}
\langle n_{k\sigma}\rangle=\frac{\exp\left(-\frac{\varepsilon_\sigma(k)}{T}\right)+\exp\left(-\frac{\varepsilon_\uparrow(k)+\varepsilon_\downarrow(k)+U}{T}\right)}{Z_k},
\end{gather}
where $Z_k=1+\exp(-\varepsilon_\uparrow(k)/T)+\exp(-\varepsilon_\downarrow(k)/T)+\exp(-(\varepsilon_\uparrow(k)+\varepsilon_\downarrow(k)+U)/T)$ is the partition sum for  mode $k$.

In the high temperature limit with 
$T\gg|U|$ and the single particle bandwidth, all occupation numbers reduce to the average occupation $\nu$, giving 
$\langle n_{k\sigma}\rangle=1/(1+\exp(-\frac{\mu}{T}))=\nu$, which can be used to determine the temperature dependence of 
the chemical potential as $\mu=T\ln\left(\nu/\left(1-\nu\right)\right)$ to leading order in $T$.
In this limit, the Green's function simplifies to
\begin{gather}
G_\sigma(k,\omega_n)=\frac{1-\nu}{i\omega_n-\varepsilon_\sigma(k)}+\frac{\nu}{i\omega_n-\varepsilon_\sigma(k)-U},
\label{eqgreen}
\end{gather}
where we still keep the Matsubara frequencies. The numerators are the same for both spin species at infinite temperatures, but the poles can be distinct when moving away from the symmetric, $\varepsilon_\uparrow(k)=\varepsilon_\downarrow(k)$ limit.
This \emph{interacting} Green's function is the weighted average  of two \emph{non-interacting} Green's function, facilitating analytical progress.
Using this, any  correlation function of interest can be obtained using standard diagrammatic rules\cite{mahan}.

The interacting case amounts to the self-energy as 
$\Sigma_\sigma(k,\omega_n)=\nu U+\frac{\nu(1-\nu)U^2}{i\omega_n-\varepsilon_\sigma(k)-(1-\nu)U}$ in the high temperature limit, 
where the first term represents the Hartree contribution. The self energy diverges even at infinite temperatures, indicating the system  cannot be continued adiabatically to the non-interacting Fermi gas but rather forms a non-Fermi liquid.
This form is also a common assumption for the self-energy deep in the Mott phase\cite{wagner,blason}.
Interestingly, the pole structure of this self energy agrees with the second order ring diagram self energy of the one dimensional Hubbard model at 
infinite temperature, which is $\Sigma(k,\omega_n)= \sum_{p,q}    \frac{U^2\nu(1-\nu)}{i\omega_n-\varepsilon(p)+\varepsilon(p-q)-\varepsilon(k-q)}$ for the symmetric case.
It diverges for $i\omega_n=\varepsilon(k)$.

These suggest that our results may find application beyond the family of Hatsugai-Kohmoto models. 
 Eq.~\eqref{eqgreen} contains all self energy correction exactly from the interaction $U$, while finite momentum vertex corrections vanish due to the locality of the interaction in 
momentum space when calculating response functions\cite{nogueira,Phillips2018,morimoto}.
Then, the spin or charge correlation functions are obtained as
\begin{gather}
\chi(q,\omega_m)=-\frac{T}{V}\sum_{nkss'}A_{ss'}(k)G_{s'}(k,\omega_n)\times\nonumber\\
\times B_{s's}(k+q)G_s(k+q,\omega_n+\omega_m),
\end{gather}
where $V$ is the system volume and $\omega_m=2m\pi T$ is the bosonic Matsubara frequency with $m$ integer. Here the $A$ and $B$ represents the matrix elements of operators $\hat A$ and $\hat B$ between the spin $s$ and $s'$ states. 
In particular, for the density correlator, these are $\delta_{s,s'}$,
while for the spin response, there are the elements of the Pauli matrices as  $\sigma^j_{s,s'}$ with $j=x,y,z$.

\section{Linear response at high temperature}

\begin{figure}[ht]
\centering
\includegraphics[width=8cm]{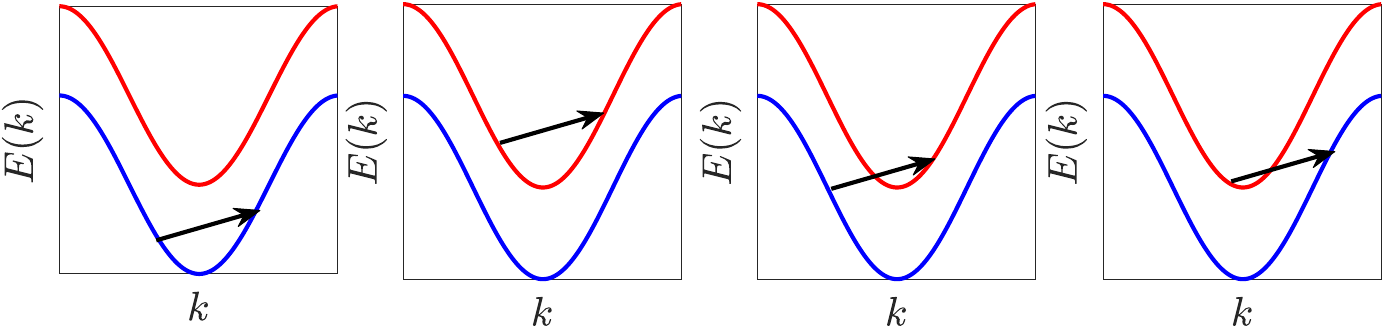}
\caption{The possible transitions for finite $q$ and $\omega$ are visualized schematically between various bands. The blue and red lines denote the lower, $\varepsilon_\sigma(k)$ 
and upper, $\varepsilon_{\sigma'}(k)+U$ Hubbard band, respectively.
With increasing $U$, the last transition becomes impossible due to energy conservation when a clean gap appears between these band. 
Further increasing the interaction pushes the third process to very high energies, therefore
in the large $U$ limit, only the first 2 transitions are possible, characterized by $w_0$.}
\label{transitions}
\end{figure}

Among the possible correlators, we are interested in the longitudinal, $\langle S_z(x,t)S_z(0,0)\rangle $ correlator, which has the same structure as the 
density-density correlation function, and the transverse or spin flip correlation function, $\langle S^+(x,t)S^-(,0,)\rangle $.
The corresponding Kubo formula is evaluated after Matsubara summation and analytic continuation to real frequencies ($i\omega_m\rightarrow\omega +i\delta$, $\delta\rightarrow 0^+$) as
\begin{gather}
\chi_{zz}(q,\omega)=\sum_\sigma\left(w_0\Phi_{\sigma\sigma}(q,\omega,0)+\sum_{\lambda=\pm }w_1\Phi_{\sigma\sigma}(q,\omega,\lambda U)\right),\\
\chi_{+-}(q,\omega)=w_0\Phi_{\uparrow\downarrow}(q,\omega,0)+\sum_{\lambda=\pm}w_1\Phi_{\uparrow\downarrow}(q,\omega,\lambda U),
\end{gather}
where
$w_0=(1-\nu)^2+\nu^2$ denotes the weight for transitions between the same Hubbard band $\varepsilon_\sigma(k)\rightarrow \varepsilon_{\sigma'}(k')$
or $\varepsilon_\sigma(k)+U\rightarrow\varepsilon_{\sigma'}(k')+U$ while $w_1=\nu(1-\nu)$ stands for transitions connecting different Hubbard bands, i.e. 
$\varepsilon_\sigma(k)\rightarrow\varepsilon_{\sigma'}(k')+U$ or $\varepsilon_\sigma(k)+U\rightarrow\varepsilon_{\sigma'}(k')$. These satisfy $w_0+2w_1=1$, $w_0\geq 2w_1$ and are only
equal for half filling, $\nu=1/2$. These transitions
are visualized in Fig. \ref{transitions}.

We defined
\begin{gather}
\Phi_{\sigma\sigma'}(q,\omega,U)=\frac{\nu(1-\nu)}{TV} \times\nonumber\\
\times\sum_{k}
\frac{\varepsilon_{\sigma'}(k)-\varepsilon_\sigma(k+q)+U}{\omega+i\delta-\varepsilon_\sigma(k+q)+\varepsilon_{\sigma'}(k)+U}.
\label{res0}
\end{gather}
This expression is obtained in the high temperature limit of the Lindhard function\cite{mahan,klasszikus,Giuliani2005} after expanding the Fermi 
distribution functions up to first order in $1/T$. Since it depends on the difference of energies, the chemical potential dependence drops out 
under the momentum sum and only remains present in the occupation number dependent prefactor.
The imaginary part of the response function simplifies to
\begin{align}
\frac{\textmd{Im}\Phi_{\sigma\sigma'}(q,\omega,U)}{\nu(1-\nu)}&=\frac{\omega\pi}{TV}\sum_k \delta\left(\omega-\varepsilon_\sigma(k+q)+\varepsilon_{\sigma'}(k)+U\right)\nonumber\\
&\equiv \frac{\omega}{T}\phi_{\sigma\sigma'}(q,\omega+U).
\label{res1}
\end{align}
In order to proceed analytically, we focus on one dimension, use the tight-binding dispersions $\varepsilon_\sigma(k)=-t_\sigma\cos(ka)$, and obtain
\begin{gather}
\phi_{\sigma\sigma'}(q,\omega)=\left((t_\sigma-t_{\sigma'})^2+4t_\sigma t_{\sigma'}\sin^2(qa/2)-\omega^2\right)^{-\frac 12}
\label{squareroot}
\end{gather}
when the expression under the negative square root is positive and zero otherwise.

\begin{figure*}[th]
\centering
\includegraphics[width=18cm]{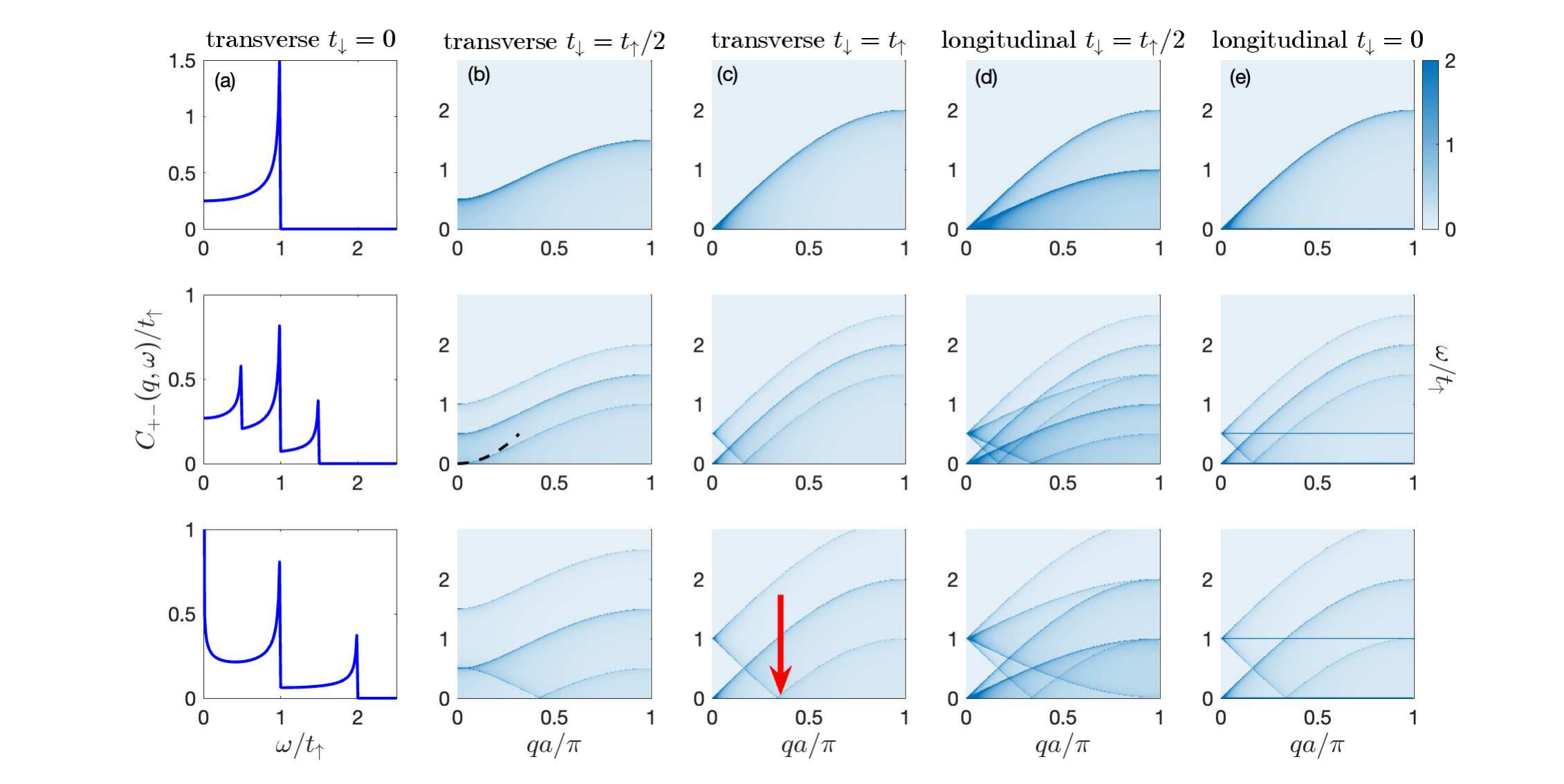}
\caption{The evolution of the transverse and longitudinal dynamical spin structure factors are visualized for half filling, $\nu=1/2$ for both spin orientations. 
The interaction varies as $U/t_\uparrow=0$, 0.5 and 1 from top to bottom and the results are insensitive to the sign of $U$..
  Column (c) shows the transverse spin structure factor 
for the symmetric case, the longitudinal one is equivalent to it up to an overall factor of 2.
Columns (b) and (d) correspond to the transverse and longitudinal spin structure factor for $t_\downarrow=t_\uparrow/2$, while
columns (a) and (e) depicts the same quantities for $t_\downarrow=0$, respectively. 
The center panel in column (b) corresponds to $t_\uparrow-t_\downarrow=U$, and displays a Lifshitz transition when the $\omega\sim |q|$ 
sound like excitation changes to $\omega=U (qa)^2$ for small $\omega$ and $q$, denoted by black dashed line.
Column (a)  corresponds to the transverse spin structure factor in the fully asymmetric case with no momentum dependence due to the flat band, 
therefore the explicit frequency dependence is displayed, revealing the characteristic inverse square root singularities from Eq.~\eqref{squareroot}.
Small energy but finite momentum excitations are also present in, e.g., the bottom panel of column (c), denoted by red arrow.
 The color encoding is the same for the 12 right panels.}
\label{dssfhk}
\end{figure*}

\section{Dynamical spin structure factor}

Using the fluctuation dissipation theorem, this allows us to obtain the dynamical spin structure factor $C_{ij}(q,\omega)$,
which is the spatial and temporal Fourier transform of the spin correlation function $\langle S^i(x,t)S^j(0,0)\rangle$, as 
\begin{gather}
C_{ij}(q,\omega)=\frac{T}{\omega}\chi^{\prime\prime}_{ij}(q,\omega)
\end{gather}
 in the high temperature ``classical'' limit and $\chi^{\prime\prime}_{ij}(q,\omega)$ denotes the imaginary part of the susceptibility.
We obtain the longitudinal and transverse spin structure factors as
\begin{subequations}
\begin{align}\label{res2}
&C_{zz}(q,\omega)=\sum_\sigma\left(\tilde w_0\phi_{\sigma\sigma}(q,\omega)+\sum_{\lambda=\pm }\tilde w_1\phi_{\sigma\sigma}(q,\omega+\lambda U)\right)\!,\\
&C_{+-}(q,\omega)=\tilde w_0\phi_{\uparrow\downarrow}(q,\omega)+\sum_{\lambda=\pm}\tilde w_1\phi_{\uparrow\downarrow}(q,\omega+\lambda U),
\end{align}
\label{C's}
\end{subequations}
respectively with $\tilde w_{0,1}=\nu(1-\nu)w_{0,1}$ carrying the occupation dependent contributions. For low ($\nu \sim 0$) or high ($\nu \sim 1$) filling factors, the $w_0$ contribution dominates
over $w_1$ and most of the spectral weight is carried by the former.

This spin structure factor reveals a square root singularity in frequency, which is expected on general grounds\cite{mullerprb} at the three dispersion branches, which mark
the boundary of the spinon continuum  
\begin{gather}
\omega=\alpha U\pm\sqrt{(t_\sigma-t_{\sigma'})^2+4t_\sigma t_{\sigma'}\sin^2(qa/2)}
 \label{spinons}
\end{gather}
with $\alpha=-1$, 0 and 1. 
For $U=0$ and $t_\uparrow=t_\downarrow$, we recover the known result corresponding to the XX Heisenberg model\cite{katsura} in the top left panel of Fig.~\ref{dssfhk}.
In contrast to $T=0$ case with an upper and lower spinon boundary, the $T=\infty$ case 
is not sensitive  to filling constraints when determining possible transitions therefore only the upper spinon boundary exists from Eq.~\eqref{spinons} 
but there is no strict lower boundary, practically
any $q$ and $\omega$ transitions are allowed below the upper boundary, similarly to the case of spin-1/2 XX chain\cite{katsura}.
By comparing the two extreme limits of $U=0$ and $U=\infty$, the ensuing spin structure factors display identical parameter dependence, though $2w_1$ fraction of
 the spectral weight is lost
in the $U=\infty$  case due to gapping some of the modes in Eq.~\eqref{res2}. The structure factor is also insensitive to the sign of $U$, as is evident from Eq. \eqref{C's}.
The maximal excitation frequency for the longitudinal spin structure factor is $|U|+2|t_\sigma|$, while it is
$|U|+|t_\uparrow|+|t_\downarrow|$ for the transverse case.

In the symmetric case with $t_\uparrow=t_\downarrow$, both longitudinal and transverse cases behave identically, as follows from SU(2) invariance.
There is a gapless branch describing sound like excitation $\omega\sim |q|$ and ballistic behaviour and two gapped dispersions for small momenta, as shown in Fig. \ref{dssfhk}. In addition, there can also be a small energy
excitation at finite wavevector due to interactions, located at $qa=\pm 2\arcsin(U/2t_\uparrow)$. This zero energy transition occurs when
a transition is created from the maximum or minimum of a given Hubbard band (momentum $\pi$ or 0) to another Hubbard band with momentum $q$.
This interaction induced softening is reminiscent to the low temperature Kohn anomaly in a variety of systems\cite{kohnprl} and is highlighted in Fig. \ref{dssfhk}.

In the asymmetric case with $t_\uparrow\neq t_\downarrow$, the longitudinal correlator retains the aforementioned behaviour with distinct parametric 
dependence of the up and down spin branches.
Therefore, two gapless and four branches are expected with distinct velocities due to the hopping asymmetries.
The transverse spin structure factor displays completely different response. In general, a gap will show up for small momenta $\sim |t_\uparrow-t_\downarrow|$ as well as 
$\sim |t_\uparrow-t_\downarrow|-U$. With increasing $U$, this gap collapses at $U=|t_\uparrow-t_\downarrow|$. At this point, the dispersion acquires a quadratic  $\omega=U (qa)^2$ behaviour, which is
reminiscent of a Lifshitz transition, depicted in Fig. \ref{dssfhk}. 
At the same time, the spectral weight of this dispersion is suppressed for low and high filling whose relative weight with respect to the $U$ independent contribution is $w_1/w_0\leq 1/2$, 
therefore it is most pronounced at around half filling.

In the Falicov-Kimball limit with $\varepsilon_\downarrow(k)=0$, all momentum dependence drops out from the spin structure factor, which is a characteristic feature of
transitions involving a flat band at infinite temperature, and depends only on the 
energy with the characteristic inverse square root singularity from Eq.~\eqref{squareroot}. The longitudinal component develops a sharp Dirac-delta peak at zero frequency and 
at  $\pm U$ due to the dispersionless flat band as a signature of the Mott transition 
on top of the continuum from the propagating band.

\section{Total spin structure factor}

The total spin structure factor is the equal time spin correlation function, which follows from
\begin{gather}
C(q)=\int d\omega ~ C(q,\omega).
\end{gather}
We obtain $C_{zz}(q)=2\nu(1-\nu)=2C_{+-}(q)$, which is independent of momentum and model parameters, and only cares about the filled initial and empty final state probabilities at infinite temperature, 
namely $\nu$ and $1-\nu$ and the factor of 2 for the longitudinal component
counts the number of spin species.

\section{Discussion}

We studied the infinite temperature spin dynamics of the asymmetric or mass-imbalanced Hatsugai-Kohmoto. Our results for the dynamical spins structure factor in
Eqs.~\eqref{res0}, \eqref{res1} and \eqref{res2} are valid in arbitrary dimension, filling, band asymmetry or interaction, and allow for analytical evaluation in one dimension.
We find rich structures in the longitudinal and transverse spin structure factor, including sound like and gapped excitation modes. The transverse component reveals a Lifshitz-like transition, when the low energy excitations disperse quadratically with the momentum,
highlighting the intricate interplay of hopping asymmetry and interactions. The asymmetric, Falicov-Kimball limit is strongly influenced by the presence of the flat band, such as  
momentum independent transverse spin structure factor due to transitions involving the flat band as well as sharp Dirac-delta peaks at zero and $\pm U$ energies in its longitudinal 
counterpart.

Although the symmetric Hatsugai-Kohmoto model represents a minimal many-body, integrable,  SU(2) symmetric system, it does not display superdiffusive behaviour\cite{ljubotina}.
Our exact results represent the rare instance when infinite temperature dynamics of strongly correlated quantum systems can be analytically revealed.

\begin{acknowledgments}
This work was supported by the National Research, Development and Innovation Office - NKFIH  Project Nos. K134437 and K142179, by a grant of the Ministry of Research, Innovation and
 Digitization, CNCS/CCCDI-UEFISCDI, under projects number
PN-IV-P1-PCE-2023-0159 and PN-IV-P1-PCE-2023-0987, and by the ``Nucleu'' Program within the PNCDI 2022-2027, 
carried out with the support of MEC, project no.~27N/03.01.2023, component project code PN 23 24 01 04.
This work was also supported by the HUN-REN Hungarian Research Network through the Supported Research Groups
Programme, HUN-REN-BME-BCE Quantum Technology Research Group (TKCS-2024/34).

\end{acknowledgments}

\bibliographystyle{apsrev}
\bibliography{hk,refgraph}

\begin{thebibliography}{10}
\expandafter\ifx\csname bibnamefont\endcsname\relax
  \def\bibnamefont#1{#1}\fi
\expandafter\ifx\csname bibfnamefont\endcsname\relax
  \def\bibfnamefont#1{#1}\fi
\expandafter\ifx\csname url\endcsname\relax
  \def\url#1{\texttt{#1}}\fi
\expandafter\ifx\csname urlprefix\endcsname\relax\def\urlprefix{URL }\fi
\providecommand{\bibinfo}[2]{#2}
\providecommand{\eprint}[2][]{\url{#2}}

\bibitem{polkovnikovrmp}
\bibinfo{author}{\bibfnamefont{A.}~\bibnamefont{Polkovnikov}},
  \bibinfo{author}{\bibfnamefont{K.}~\bibnamefont{Sengupta}},
  \bibinfo{author}{\bibfnamefont{A.}~\bibnamefont{Silva}}, \bibnamefont{and}
  \bibinfo{author}{\bibfnamefont{M.}~\bibnamefont{Vengalattore}},
  \emph{\bibinfo{title}{\textit{Colloquium} : Nonequilibrium dynamics of closed
  interacting quantum systems}}, \bibinfo{journal}{Rev. Mod. Phys.}
  \textbf{\bibinfo{volume}{83}}, \bibinfo{pages}{863} (\bibinfo{year}{2011}).

\bibitem{dziarmagareview}
\bibinfo{author}{\bibfnamefont{J.}~\bibnamefont{Dziarmaga}},
  \emph{\bibinfo{title}{Dynamics of a quantum phase transition and relaxation
  to a steady state}}, \bibinfo{journal}{Adv. Phys.}
  \textbf{\bibinfo{volume}{59}}, \bibinfo{pages}{1063} (\bibinfo{year}{2010}).

\bibitem{nowak}
\bibinfo{author}{\bibfnamefont{U.}~\bibnamefont{Nowak}}, in
  \emph{\bibinfo{booktitle}{Micromagnetism}}, edited by
  \bibinfo{editor}{\bibfnamefont{H.}~\bibnamefont{Kronmüller}}
  (\bibinfo{publisher}{Wiley}, \bibinfo{address}{Chichester},
  \bibinfo{year}{2007}), no.~\bibinfo{number}{2} in \bibinfo{series}{Handbook
  of magnetism and advanced magnetic materials}, pp. \bibinfo{pages}{858--876}.

\bibitem{blundell}
\bibinfo{author}{\bibfnamefont{S.}~\bibnamefont{Blundell}},
  \emph{\bibinfo{title}{Magnetism in Condensed Matter}}, Oxford Master Series
  in Condensed Matter Physics (\bibinfo{publisher}{OUP Oxford},
  \bibinfo{year}{2001}).

\bibitem{coldea}
\bibinfo{author}{\bibfnamefont{R.}~\bibnamefont{Coldea}},
  \bibinfo{author}{\bibfnamefont{D.~A.} \bibnamefont{Tennant}},
  \bibinfo{author}{\bibfnamefont{E.~M.} \bibnamefont{Wheeler}},
  \bibinfo{author}{\bibfnamefont{E.}~\bibnamefont{Wawrzynska}},
  \bibinfo{author}{\bibfnamefont{D.}~\bibnamefont{Prabhakaran}},
  \bibinfo{author}{\bibfnamefont{M.}~\bibnamefont{Telling}},
  \bibinfo{author}{\bibfnamefont{K.}~\bibnamefont{Habicht}},
  \bibinfo{author}{\bibfnamefont{P.}~\bibnamefont{Smeibidl}}, \bibnamefont{and}
  \bibinfo{author}{\bibfnamefont{K.}~\bibnamefont{Kiefer}},
  \emph{\bibinfo{title}{Quantum criticality in an ising chain: Experimental
  evidence for emergent ${E}_8$ symmetry}}, \bibinfo{journal}{Science}
  \textbf{\bibinfo{volume}{327}}(\bibinfo{number}{5962}), \bibinfo{pages}{177}
  (\bibinfo{year}{2010}).

\bibitem{niemijer}
\bibinfo{author}{\bibfnamefont{T.}~\bibnamefont{Niemeijer}},
  \emph{\bibinfo{title}{Some exact calculations on a chain of spins 12}},
  \bibinfo{journal}{Physica}
  \textbf{\bibinfo{volume}{36}}(\bibinfo{number}{3}), \bibinfo{pages}{377}
  (\bibinfo{year}{1967}).

\bibitem{sur}
\bibinfo{author}{\bibfnamefont{A.}~\bibnamefont{Sur}},
  \bibinfo{author}{\bibfnamefont{D.}~\bibnamefont{Jasnow}}, \bibnamefont{and}
  \bibinfo{author}{\bibfnamefont{I.~J.} \bibnamefont{Lowe}},
  \emph{\bibinfo{title}{Spin dynamics for the one-dimensional $\mathrm{XY}$
  model at infinite temperature}}, \bibinfo{journal}{Phys. Rev. B}
  \textbf{\bibinfo{volume}{12}}, \bibinfo{pages}{3845} (\bibinfo{year}{1975}).

\bibitem{gerling}
\bibinfo{author}{\bibfnamefont{R.~W.} \bibnamefont{Gerling}} \bibnamefont{and}
  \bibinfo{author}{\bibfnamefont{D.~P.} \bibnamefont{Landau}},
  \emph{\bibinfo{title}{Time-dependent behavior of classical spin chains at
  infinite temperature}}, \bibinfo{journal}{Phys. Rev. B}
  \textbf{\bibinfo{volume}{42}}, \bibinfo{pages}{8214} (\bibinfo{year}{1990}).

\bibitem{katsura}
\bibinfo{author}{\bibfnamefont{S.}~\bibnamefont{Katsura}},
  \bibinfo{author}{\bibfnamefont{T.}~\bibnamefont{Horiguchi}},
  \bibnamefont{and} \bibinfo{author}{\bibfnamefont{M.}~\bibnamefont{Suzuki}},
  \emph{\bibinfo{title}{Dynamical properties of the isotropic xy model}},
  \bibinfo{journal}{Physica}
  \textbf{\bibinfo{volume}{46}}(\bibinfo{number}{1}), \bibinfo{pages}{67}
  (\bibinfo{year}{1970}).

\bibitem{ljubotina}
\bibinfo{author}{\bibfnamefont{M.}~\bibnamefont{Ljubotina}},
  \bibinfo{author}{\bibfnamefont{M.}~\bibnamefont{Znidaric}}, \bibnamefont{and}
  \bibinfo{author}{\bibfnamefont{T.}~\bibnamefont{Prosen}},
  \emph{\bibinfo{title}{Spin diffusion from an inhomogeneous quench in an
  integrable system}}, \bibinfo{journal}{Nature Communications}
  \textbf{\bibinfo{volume}{8}}, \bibinfo{pages}{16117} (\bibinfo{year}{2017}).

\bibitem{sanchez}
\bibinfo{author}{\bibfnamefont{R.~J.} \bibnamefont{S\'anchez}},
  \bibinfo{author}{\bibfnamefont{V.~K.} \bibnamefont{Varma}}, \bibnamefont{and}
  \bibinfo{author}{\bibfnamefont{V.}~\bibnamefont{Oganesyan}},
  \emph{\bibinfo{title}{Anomalous and regular transport in spin-$\frac{1}{2}$
  chains: ac conductivity}}, \bibinfo{journal}{Phys. Rev. B}
  \textbf{\bibinfo{volume}{98}}, \bibinfo{pages}{054415}
  (\bibinfo{year}{2018}).

\bibitem{gopalakrishnan}
\bibinfo{author}{\bibfnamefont{S.}~\bibnamefont{Gopalakrishnan}},
  \bibinfo{author}{\bibfnamefont{R.}~\bibnamefont{Vasseur}}, \bibnamefont{and}
  \bibinfo{author}{\bibfnamefont{B.}~\bibnamefont{Ware}},
  \emph{\bibinfo{title}{Anomalous relaxation and the high-temperature structure
  factor of xxz spin chains}}, \bibinfo{journal}{Proceedings of the National
  Academy of Sciences} \textbf{\bibinfo{volume}{116}}(\bibinfo{number}{33}),
  \bibinfo{pages}{16250} (\bibinfo{year}{2019}).

\bibitem{luitz}
\bibinfo{author}{\bibfnamefont{D.~J.} \bibnamefont{Luitz}},
  \bibinfo{author}{\bibfnamefont{R.}~\bibnamefont{Moessner}},
  \bibinfo{author}{\bibfnamefont{S.~L.} \bibnamefont{Sondhi}},
  \bibnamefont{and} \bibinfo{author}{\bibfnamefont{V.}~\bibnamefont{Khemani}},
  \emph{\bibinfo{title}{Prethermalization without temperature}},
  \bibinfo{journal}{Phys. Rev. X} \textbf{\bibinfo{volume}{10}},
  \bibinfo{pages}{021046} (\bibinfo{year}{2020}).

\bibitem{hatsugai1992}
\bibinfo{author}{\bibfnamefont{Y.}~\bibnamefont{Hatsugai}} \bibnamefont{and}
  \bibinfo{author}{\bibfnamefont{M.}~\bibnamefont{Kohmoto}},
  \emph{\bibinfo{title}{Exactly solvable model of correlated lattice electrons
  in any dimensions}}, \bibinfo{journal}{Journal of the Physical Society of
  Japan} \textbf{\bibinfo{volume}{61}}(\bibinfo{number}{6}),
  \bibinfo{pages}{2056} (\bibinfo{year}{1992}).

\bibitem{Sun2024}
\bibinfo{author}{\bibfnamefont{Z.}~\bibnamefont{Sun}} \bibnamefont{and}
  \bibinfo{author}{\bibfnamefont{H.-Q.} \bibnamefont{Lin}},
  \emph{\bibinfo{title}{Emergence of a diverse array of phases in an exactly
  solvable model}}, \bibinfo{journal}{Phys. Rev. B}
  \textbf{\bibinfo{volume}{109}}, \bibinfo{pages}{115108}
  (\bibinfo{year}{2024}).

\bibitem{zaanen}
\bibinfo{author}{\bibfnamefont{J.}~\bibnamefont{Zaanen}},
  \emph{\bibinfo{title}{Carriers that count}}, \bibinfo{journal}{Nat. Phys.}
  \textbf{\bibinfo{volume}{16}}, \bibinfo{pages}{1171} (\bibinfo{year}{2020}).

\bibitem{nesselrodt}
\bibinfo{author}{\bibfnamefont{R.~D.} \bibnamefont{Nesselrodt}}
  \bibnamefont{and} \bibinfo{author}{\bibfnamefont{J.~K.}
  \bibnamefont{Freericks}}, \emph{\bibinfo{title}{Exact solution of two simple
  non-equilibrium electron-phonon and electron-electron coupled systems: The
  atomic limit of the holstein-hubbard model and the generalized
  hatsugai-komoto model}}, \bibinfo{journal}{Phys. Rev. B}
  \textbf{\bibinfo{volume}{104}}, \bibinfo{pages}{155104}
  (\bibinfo{year}{2021}).

\bibitem{zhao2022}
\bibinfo{author}{\bibfnamefont{J.}~\bibnamefont{Zhao}},
  \bibinfo{author}{\bibfnamefont{L.}~\bibnamefont{Yeo}},
  \bibinfo{author}{\bibfnamefont{E.~W.} \bibnamefont{Huang}}, \bibnamefont{and}
  \bibinfo{author}{\bibfnamefont{P.~W.} \bibnamefont{Phillips}},
  \emph{\bibinfo{title}{Thermodynamics of an exactly solvable model for
  superconductivity in a doped mott insulator}}, \bibinfo{journal}{Phys. Rev.
  B} \textbf{\bibinfo{volume}{105}}, \bibinfo{pages}{184509}
  (\bibinfo{year}{2022}).

\bibitem{lidsky}
\bibinfo{author}{\bibfnamefont{D.}~\bibnamefont{Lidsky}},
  \bibinfo{author}{\bibfnamefont{J.}~\bibnamefont{Shiraishi}},
  \bibinfo{author}{\bibfnamefont{Y.}~\bibnamefont{Hatsugai}}, \bibnamefont{and}
  \bibinfo{author}{\bibfnamefont{M.}~\bibnamefont{Kohmoto}},
  \emph{\bibinfo{title}{Simple exactly solvable models of non-fermi-liquids}},
  \bibinfo{journal}{Phys. Rev. B} \textbf{\bibinfo{volume}{57}},
  \bibinfo{pages}{1340} (\bibinfo{year}{1998}).

\bibitem{zhu2021}
\bibinfo{author}{\bibfnamefont{H.-S.} \bibnamefont{Zhu}} \bibnamefont{and}
  \bibinfo{author}{\bibfnamefont{Q.}~\bibnamefont{Han}},
  \emph{\bibinfo{title}{Effects of electron correlation on superconductivity in
  the hatsugai–kohmoto model*}}, \bibinfo{journal}{Chinese Physics B}
  \textbf{\bibinfo{volume}{30}}(\bibinfo{number}{10}), \bibinfo{pages}{107401}
  (\bibinfo{year}{2021}).

\bibitem{zhao2025}
\bibinfo{author}{\bibfnamefont{M.}~\bibnamefont{Zhao}},
  \bibinfo{author}{\bibfnamefont{W.-W.} \bibnamefont{Yang}}, \bibnamefont{and}
  \bibinfo{author}{\bibfnamefont{Y.}~\bibnamefont{Zhong}},
  \emph{\bibinfo{title}{Hatsugai–kohmoto models: exactly solvable playground
  for mottness and non-fermi liquid}}, \bibinfo{journal}{Journal of Physics:
  Condensed Matter} \textbf{\bibinfo{volume}{37}}(\bibinfo{number}{18}),
  \bibinfo{pages}{183005} (\bibinfo{year}{2025}).

\bibitem{skolimowski}
\bibinfo{author}{\bibfnamefont{J.}~\bibnamefont{Skolimowski}},
  \emph{\bibinfo{title}{Real-space analysis of hatsugai-kohmoto interaction}},
  \bibinfo{journal}{Phys. Rev. B} \textbf{\bibinfo{volume}{109}},
  \bibinfo{pages}{165129} (\bibinfo{year}{2024}).

\bibitem{morimoto}
\bibinfo{author}{\bibfnamefont{T.}~\bibnamefont{Morimoto}} \bibnamefont{and}
  \bibinfo{author}{\bibfnamefont{N.}~\bibnamefont{Nagaosa}},
  \emph{\bibinfo{title}{Weyl mott insulator}}, \bibinfo{journal}{Scientific
  Reports} \textbf{\bibinfo{volume}{6}}, \bibinfo{pages}{19853}
  (\bibinfo{year}{2016}).

\bibitem{Phillips2020}
\bibinfo{author}{\bibfnamefont{P.~W.} \bibnamefont{Phillips}},
  \bibinfo{author}{\bibfnamefont{L.}~\bibnamefont{Yeo}}, \bibnamefont{and}
  \bibinfo{author}{\bibfnamefont{E.~W.} \bibnamefont{Huang}},
  \emph{\bibinfo{title}{Exact theory for superconductivity in a doped mott
  insulator}}, \bibinfo{journal}{Nature Physics}
  \textbf{\bibinfo{volume}{16}}(\bibinfo{number}{12}), \bibinfo{pages}{1175}
  (\bibinfo{year}{2020}).

\bibitem{falicov}
\bibinfo{author}{\bibfnamefont{L.~M.} \bibnamefont{Falicov}} \bibnamefont{and}
  \bibinfo{author}{\bibfnamefont{J.~C.} \bibnamefont{Kimball}},
  \emph{\bibinfo{title}{Simple model for semiconductor-metal transitions:
  Sm${\mathrm{b}}_{6}$ and transition-metal oxides}}, \bibinfo{journal}{Phys.
  Rev. Lett.} \textbf{\bibinfo{volume}{22}}, \bibinfo{pages}{997}
  (\bibinfo{year}{1969}).

\bibitem{baez}
\bibinfo{author}{\bibfnamefont{M.~L.} \bibnamefont{Baez}},
  \bibinfo{author}{\bibfnamefont{M.}~\bibnamefont{Goihl}},
  \bibinfo{author}{\bibfnamefont{J.}~\bibnamefont{Haferkamp}},
  \bibinfo{author}{\bibfnamefont{J.}~\bibnamefont{Bermejo-Vega}},
  \bibinfo{author}{\bibfnamefont{M.}~\bibnamefont{Gluza}}, \bibnamefont{and}
  \bibinfo{author}{\bibfnamefont{J.}~\bibnamefont{Eisert}},
  \emph{\bibinfo{title}{Dynamical structure factors of dynamical quantum
  simulators}}, \bibinfo{journal}{Proceedings of the National Academy of
  Sciences} \textbf{\bibinfo{volume}{117}}(\bibinfo{number}{42}),
  \bibinfo{pages}{26123} (\bibinfo{year}{2020}).

\bibitem{Farkasovsky}
\bibinfo{author}{\bibfnamefont{P.}~\bibnamefont{Farkasovsk\'y}},
  \emph{\bibinfo{title}{Phase diagram of the asymmetric hubbard model}},
  \bibinfo{journal}{Phys. Rev. B} \textbf{\bibinfo{volume}{77}},
  \bibinfo{pages}{085110} (\bibinfo{year}{2008}).

\bibitem{maska}
\bibinfo{author}{\bibfnamefont{M.~M.} \bibnamefont{Maska}} \bibnamefont{and}
  \bibinfo{author}{\bibfnamefont{K.}~\bibnamefont{Czajka}},
  \emph{\bibinfo{title}{Thermodynamics of the two-dimensional falicov-kimball
  model: A classical monte carlo study}}, \bibinfo{journal}{Phys. Rev. B}
  \textbf{\bibinfo{volume}{74}}, \bibinfo{pages}{035109}
  (\bibinfo{year}{2006}).

\bibitem{wang2007}
\bibinfo{author}{\bibfnamefont{Z.~G.} \bibnamefont{Wang}},
  \bibinfo{author}{\bibfnamefont{Y.~G.} \bibnamefont{Chen}}, \bibnamefont{and}
  \bibinfo{author}{\bibfnamefont{S.~J.} \bibnamefont{Gu}},
  \emph{\bibinfo{title}{Bosonization study of quantum phase transitions in the
  one-dimensional asymmetric hubbard model}}, \bibinfo{journal}{Phys. Rev. B}
  \textbf{\bibinfo{volume}{75}}, \bibinfo{pages}{165111}
  (\bibinfo{year}{2007}).

\bibitem{heitmann}
\bibinfo{author}{\bibfnamefont{T.}~\bibnamefont{Heitmann}},
  \bibinfo{author}{\bibfnamefont{J.}~\bibnamefont{Richter}},
  \bibinfo{author}{\bibfnamefont{T.}~\bibnamefont{Dahm}}, \bibnamefont{and}
  \bibinfo{author}{\bibfnamefont{R.}~\bibnamefont{Steinigeweg}},
  \emph{\bibinfo{title}{Density dynamics in the mass-imbalanced hubbard
  chain}}, \bibinfo{journal}{Phys. Rev. B} \textbf{\bibinfo{volume}{102}},
  \bibinfo{pages}{045137} (\bibinfo{year}{2020}).

\bibitem{jin2015}
\bibinfo{author}{\bibfnamefont{F.}~\bibnamefont{Jin}},
  \bibinfo{author}{\bibfnamefont{R.}~\bibnamefont{Steinigeweg}},
  \bibinfo{author}{\bibfnamefont{F.}~\bibnamefont{Heidrich-Meisner}},
  \bibinfo{author}{\bibfnamefont{K.}~\bibnamefont{Michielsen}},
  \bibnamefont{and} \bibinfo{author}{\bibfnamefont{H.}~\bibnamefont{De~Raedt}},
  \emph{\bibinfo{title}{Finite-temperature charge transport in the
  one-dimensional hubbard model}}, \bibinfo{journal}{Phys. Rev. B}
  \textbf{\bibinfo{volume}{92}}, \bibinfo{pages}{205103}
  (\bibinfo{year}{2015}).

\bibitem{zlatic}
\bibinfo{author}{\bibfnamefont{J.~K.} \bibnamefont{Freericks}}
  \bibnamefont{and} \bibinfo{author}{\bibfnamefont{V.}~\bibnamefont{Zlatic}},
  \emph{\bibinfo{title}{Exact dynamical mean-field theory of the
  falicov-kimball model}}, \bibinfo{journal}{Rev. Mod. Phys.}
  \textbf{\bibinfo{volume}{75}}, \bibinfo{pages}{1333} (\bibinfo{year}{2003}).

\bibitem{freericks}
\bibinfo{author}{\bibfnamefont{J.}~\bibnamefont{Freericks}},
  \bibinfo{author}{\bibfnamefont{E.}~\bibnamefont{Lieb}}, \bibnamefont{and}
  \bibinfo{author}{\bibfnamefont{D.}~\bibnamefont{Ueltschi}},
  \emph{\bibinfo{title}{Segregation in the falicov--kimball model}},
  \bibinfo{journal}{Commun. Math. Phys.} \textbf{\bibinfo{volume}{227}},
  \bibinfo{pages}{243} (\bibinfo{year}{2002}).

\bibitem{Li2019}
\bibinfo{author}{\bibfnamefont{X.-H.} \bibnamefont{Li}},
  \bibinfo{author}{\bibfnamefont{Z.}~\bibnamefont{Chen}}, \bibnamefont{and}
  \bibinfo{author}{\bibfnamefont{T.~K.} \bibnamefont{Ng}},
  \emph{\bibinfo{title}{Generalized falicov-kimball models}},
  \bibinfo{journal}{Phys. Rev. B} \textbf{\bibinfo{volume}{100}},
  \bibinfo{pages}{094519} (\bibinfo{year}{2019}).

\bibitem{arovas}
\bibinfo{author}{\bibfnamefont{D.~P.} \bibnamefont{Arovas}},
  \bibinfo{author}{\bibfnamefont{E.}~\bibnamefont{Berg}},
  \bibinfo{author}{\bibfnamefont{S.~A.} \bibnamefont{Kivelson}},
  \bibnamefont{and} \bibinfo{author}{\bibfnamefont{S.}~\bibnamefont{Raghu}},
  \emph{\bibinfo{title}{The hubbard model}}, \bibinfo{journal}{Annual Review of
  Condensed Matter Physics} \textbf{\bibinfo{volume}{13}}(\bibinfo{number}{1}),
  \bibinfo{pages}{239} (\bibinfo{year}{2022}).

\bibitem{mahan}
\bibinfo{author}{\bibfnamefont{G.~D.} \bibnamefont{Mahan}},
  \emph{\bibinfo{title}{Many particle physics}} (\bibinfo{publisher}{Plenum
  Publishers}, \bibinfo{address}{New York}, \bibinfo{year}{1990}).

\bibitem{wagner}
\bibinfo{author}{\bibfnamefont{N.}~\bibnamefont{Wagner}},
  \bibinfo{author}{\bibfnamefont{L.}~\bibnamefont{Crippa}},
  \bibinfo{author}{\bibfnamefont{A.}~\bibnamefont{Amaricci}},
  \bibinfo{author}{\bibfnamefont{P.}~\bibnamefont{Hansmann}},
  \bibinfo{author}{\bibfnamefont{M.}~\bibnamefont{Klett}},
  \bibinfo{author}{\bibfnamefont{E.~J.} \bibnamefont{K{\"o}nig}},
  \bibinfo{author}{\bibfnamefont{T.}~\bibnamefont{Schafer}},
  \bibinfo{author}{\bibfnamefont{D.~D.} \bibnamefont{Sante}},
  \bibinfo{author}{\bibfnamefont{J.}~\bibnamefont{Cano}},
  \bibinfo{author}{\bibfnamefont{A.~J.} \bibnamefont{Millis}},
  \bibinfo{author}{\bibfnamefont{A.}~\bibnamefont{Georges}}, \bibnamefont{and}
  \bibinfo{author}{\bibfnamefont{G.}~\bibnamefont{Sangiovanni}},
  \emph{\bibinfo{title}{Mott insulators with boundary zeros}},
  \bibinfo{journal}{Nature Communications} \textbf{\bibinfo{volume}{14}},
  \bibinfo{pages}{7531} (\bibinfo{year}{2023}).

\bibitem{blason}
\bibinfo{author}{\bibfnamefont{A.}~\bibnamefont{Blason}} \bibnamefont{and}
  \bibinfo{author}{\bibfnamefont{M.}~\bibnamefont{Fabrizio}},
  \emph{\bibinfo{title}{Unified role of green's function poles and zeros in
  correlated topological insulators}}, \bibinfo{journal}{Phys. Rev. B}
  \textbf{\bibinfo{volume}{108}}, \bibinfo{pages}{125115}
  (\bibinfo{year}{2023}).

\bibitem{nogueira}
\bibinfo{author}{\bibfnamefont{F.~S.} \bibnamefont{Nogueira}} \bibnamefont{and}
  \bibinfo{author}{\bibfnamefont{E.~V.} \bibnamefont{Anda}},
  \emph{\bibinfo{title}{Study on a toy model of strongly correlated
  electrons}}, \bibinfo{journal}{International Journal of Modern Physics B}
  \textbf{\bibinfo{volume}{10}}(\bibinfo{number}{27}), \bibinfo{pages}{3705}
  (\bibinfo{year}{1996}).

\bibitem{Phillips2018}
\bibinfo{author}{\bibfnamefont{P.~W.} \bibnamefont{Phillips}},
  \bibinfo{author}{\bibfnamefont{C.}~\bibnamefont{Setty}}, \bibnamefont{and}
  \bibinfo{author}{\bibfnamefont{S.}~\bibnamefont{Zhang}},
  \emph{\bibinfo{title}{Absence of a charge diffusion pole at finite energies
  in an exactly solvable interacting flat-band model in $d$ dimensions}},
  \bibinfo{journal}{Phys. Rev. B} \textbf{\bibinfo{volume}{97}},
  \bibinfo{pages}{195102} (\bibinfo{year}{2018}).

\bibitem{klasszikus}
\bibinfo{author}{\bibfnamefont{A.~A.} \bibnamefont{Abrikosov}},
  \bibinfo{author}{\bibfnamefont{L.~P.} \bibnamefont{Gor'kov}},
  \bibnamefont{and} \bibinfo{author}{\bibfnamefont{I.~E.}
  \bibnamefont{Dzyaloshinski}}, \emph{\bibinfo{title}{Methods of Quantum Field
  Theory in Statistical Physics}} (\bibinfo{publisher}{Dover Publications},
  \bibinfo{address}{New York}, \bibinfo{year}{1963}).

\bibitem{Giuliani2005}
\bibinfo{author}{\bibfnamefont{G.}~\bibnamefont{Giuliani}} \bibnamefont{and}
  \bibinfo{author}{\bibfnamefont{G.}~\bibnamefont{Vignale}},
  \emph{\bibinfo{title}{Quantum Theory of the Electron Liquid}}
  (\bibinfo{publisher}{Cambridge University Press}, \bibinfo{year}{2005}).

\bibitem{mullerprb}
\bibinfo{author}{\bibfnamefont{G.}~\bibnamefont{M\"uller}},
  \bibinfo{author}{\bibfnamefont{H.}~\bibnamefont{Thomas}},
  \bibinfo{author}{\bibfnamefont{H.}~\bibnamefont{Beck}}, \bibnamefont{and}
  \bibinfo{author}{\bibfnamefont{J.~C.} \bibnamefont{Bonner}},
  \emph{\bibinfo{title}{Quantum spin dynamics of the antiferromagnetic linear
  chain in zero and nonzero magnetic field}}, \bibinfo{journal}{Phys. Rev. B}
  \textbf{\bibinfo{volume}{24}}, \bibinfo{pages}{1429} (\bibinfo{year}{1981}).

\bibitem{kohnprl}
\bibinfo{author}{\bibfnamefont{W.}~\bibnamefont{Kohn}},
  \emph{\bibinfo{title}{Image of the fermi surface in the vibration spectrum of
  a metal}}, \bibinfo{journal}{Phys. Rev. Lett.} \textbf{\bibinfo{volume}{2}},
  \bibinfo{pages}{393} (\bibinfo{year}{1959}).

\end{thebibliography}

\end{document}